\documentclass[aps,nofootinbib,preprintnumbers,preprint]{revtex4-1}
\usepackage{geometry}     
\usepackage{natbib}
\usepackage{graphicx}
\usepackage{amsmath,amssymb}  
\usepackage{bm}  
\usepackage{slashed}
\usepackage{verbatim}
\usepackage{caption}
\usepackage{subfig}
\usepackage{psfrag}
\usepackage{psfrag}
\newcommand\beq{\begin{equation}}
\newcommand\eeq{\end{equation}}

\newcommand\nn{\nonumber}
\newcommand\vev[1]{\langle #1\rangle}

\newcommand\hc{\text{h.c.}}
\newcommand\fba{A^{t\bar{t}}_{FB}}

\begin{document}
\title{Explaining the $t\bar{t}$ Forward-Backward Asymmetry from a GUT-Inspired Model}


\author{David C. Stone}
\email[Electronic address: ]{dcstone@physics.ucsd.edu}
\affiliation{Department of Physics, University of California at San Diego, La Jolla, CA 92093}

\author{Patipan Uttayarat}
\email[Electronic address: ]{puttayarat@physics.ucsd.edu}
\affiliation{Department of Physics, University of California at San Diego, La Jolla, CA 92093}

\preprint{UCSD PTH 11-20}

\begin{abstract}
We consider a model that includes light colored scalars from the {\bf 45} and {\bf 50} representations of $SU(5)$ in order to explain the CDF- and D$\O$-reported $t\bar{t}$ forward-backward asymmetry, $A^{t\bar{t}}_{FB}$.
These light scalars are, labeled by their charges under the Standard Model gauge groups, the $(6,1)_{4/3}$ and $(\bar{6},3)_{-1/3}$ from the {\bf 50} and the $(8,2)_{1/2}$ from the {\bf 45}.
When the Yukawa coupling of the {\bf 50} is reasonably chosen and that of the {\bf 45} kept negligible at the scale of $M_Z$, the model yields phenomenologically viable results in agreement with the total $A^{t\bar{t}}_{FB}$ reported by CDF at the 0.7$\sigma$ level and with $A^{t\bar{t}}_{FB} (M_{t\bar{t}} \ge 450 \text{ GeV})$ at the 2.2$\sigma$ level.
Additionally, the Yukawa coupling of the {\bf 50} remains perturbative to the GUT scale (which defines a ``reasonably chosen'' value at $M_Z$), and the presence of the light scalar from the {\bf 45} allows for gauge coupling unification at a scale of $M_{GUT} \sim 10^{17}$ GeV.
\end{abstract}

\maketitle

\section{Introduction}

The Standard Model (SM) has been a very successful model when confronted with experimental observations. 
However, there are motivations to study New Physics (NP) that supersede or extend the SM. 
The reasons are two-fold.  
On the one hand, there are anomalies reported from various experiments that cannot be explained by the SM.  
If these anomalies are verified, they necessary imply NP.  
On the other hand, the study of NP has been fueled by theoretical curiosities.  
One oft-studied scenario is the possibility of the unification of fundamental forces.  
In this paper, we will explore a NP model that could explain reported anomalies while at the same time allowing for the unification of fundamental forces.

From the theoretical point of view, the SM suggests the three gauge forces of $SU(3)_C\times SU(2)_L\times U(1)_Y$ unify at at high scale ($\sim10^{15}$ GeV). 
This observation leads to the formulation of a grand unified theory (GUT) in which all three gauge forces originate from just one fundamental gauge group.  
The simplest such model is the minimal $SU(5)$ model of Georgi and Glashow~\cite{Georgi:1974sy}. 
However, this minimal model predicts an incorrect fermion mass ratio.  
To make the model phenomenologically viable, scalar fields transforming in the 45-dimensional representation of $SU(5)$ are introduced~\cite{Georgi:1979df}. 
This raises the possibility that there could also be more scalar fields transforming in some other representations of $SU(5)$.  
If some components of these scalar fields are light, they could be relevant for low energy physics.

On experimental side, the CDF and D$\O$ collaboration have recently reported a measurement of the $t\bar{t}$ forward-backward asymmetry ($A_{FB}^{t\bar{t}}$)~\cite{Abazov:2011rq,Aaltonen:2011kc,CDFnote:10584} which deviates from the SM prediction~\cite{Ahrens:2010zv} at more than the $2\sigma$ level.  
Moreover, CDF also reports that the asymmetry grows with the invariant mass of the $t\bar{t}$ system.  
In particular, the CDF measurement of $A_{FB}^{t\bar{t}}$ for $M_{t\bar{t}}\ge 450$ GeV~\cite{Aaltonen:2011kc} is more than $3\sigma$ away from the SM predictions~\cite{Ahrens:2010zv}. 
These discrepancies invite NP explanations. 
There are many models proposed in the literature to explain the $A_{FB}^{t\bar{t}}$ anomaly.  
Most of them involve the introduction of a new particle near the electroweak scale, see refs.~\cite{Shu:2009xf,Dorsner:2009mq,Barger:2010mw,Dorsner:2011ai,Shelton:2011hq,Nelson:2011us,Grinstein:2011yv,Grinstein:2011dz,Cheung:2009ch,Patel:2011eh} for a partial list of references.  

In this work we focus our attention on models involving new colored scalar fields. 
This class of models is interesting in our opinion since the scalar fields could arise from GUT scalar multiplets. 
The model with an extra scalar field in various representations has been previously studied in ref.~\cite{Shu:2009xf}.  
However, to generate a large $A_{FB}^{t\bar{t}}$ consistent with CDF and D$\O$ measurements, the scalar Yukawa couplings are generally taken to be large.  
Such a large Yukawa coupling would become non-perturbative at a scale not far above the weak scale. 
This difficulty can be overcome by having multiple light scalars contribute to the $A_{FB}^{t\bar{t}}$.   
As an added benefit, multiple light scalar fields can conspire to give gauge coupling unification.  
This idea has been previously explored by Dorsner et. al.~\cite{Dorsner:2009mq, Dorsner:2011ai}.  
However, the scalar field studied by Dorsner et. al. can couple quarks to leptons and mediate proton decay via a dimension-9 operator.
Bounds on proton decay lead to a lower bound on the mass of their scalar of $\sim 10^{10}$ GeV, far too high to be of relevance to $t\bar{t}$ phenomenology.
Thus we seek different scalar representations which could unify gauge couplings, explain the $A_{FB}^{t\bar{t}}$, and not lead to proton decay.
Previous papers, in particular a model with multiple light colored scalars from an $SO(10)$ GUT, have had some success with this approach~\cite{Patel:2011eh}.
However, those results introduced scalars with masses at both the electroweak scale and at an intermediate scale of approximately $10^9$-$10^{12}$ GeV. 
In our model, it is not necessary to introduce any scales other than the electroweak and GUT scale; we consider this a distinct advantage.

One might object that adding light scalars leads to a hierarchy problem.  
In fact, even in the minimal $SU(5)$ model the mass of the Higgs doublet has to be fine tuned to be at the electroweak scale.   
We will assume here that the hierarchy problem can be solved in a similar fashion, and that a similar mechanism is responsible for masses of the new scalar fields at the electroweak scale.

The structure of this paper is as follows: In section~\ref{sec:couplingunification} we discuss the effects these colored scalars have on gauge coupling unification.
In section~\ref{sec:phenom} we discuss the effects these particles have on general $t\bar{t}$ phenomenology, including the total cross-section and $A_{FB}^{t\bar{t}}$.
Finally, in section~\ref{sec:conclusions} we draw general conclusions of the merits of this model, its weaknesses, and look towards the LHC phenomenology of the model.

\section{Gauge Coupling Unification}
\label{sec:couplingunification}
Upon closer inspection, the three gauge couplings of the SM do not quite unify.  
However, if one allows for more fields at low energy, the unification of gauge coupling could be achieved. To have a consistent GUT, these light particles must be part of some incomplete representation of the GUT gauge group.  
For definiteness, we will consider GUT model based on the $SU(5)$ gauge group.  
We will first give a brief review of the $SU(5)$ GUT model.
  
In the minimal $SU(5)$ model, each family of the SM fermion content is embedded in the {\bf 5} and the {\bf 10} representation of $SU(5)$ as follows
\begin{equation}
\begin{split}
	\chi_R &= \begin{pmatrix}d_{R_1}\\d_{R_2}\\d_{R_3}\\ \overline{e_L} \\ -\overline{\nu_L}\end{pmatrix},\quad
	\Psi_L = \begin{pmatrix}
		0& \overline{u_R}_3 &-\overline{u_R}_2 & -u_{L_1} & -d_{L_1}\\ 
		-\overline{u_R}_3 & 0 &\overline{u_R}_1 & -u_{L_2} & -d_{L_2}\\
		 \overline{u_R}_2 &-\overline{u_R}_1 & 0 & -u_{L_3} & -d_{L_3} \\
		 u_{L_1} & u_{L_2} & u_{L_3} & 0 & \overline{e_R}\\
		 d_{L_1} & d_{L_2} & d_{L_3} & -\overline{e_R} & 0
	\end{pmatrix},
\end{split}
\end{equation}
where we use the convention that the {\bf 5} of $SU(5)$ decomposes to $(3,1)_{-1/3} \oplus (1,2)_{1/2}$ under the SM gauge group.  In the minimal setup, there are two scalar representations, the {\bf 5} and the {\bf 24}, denoted by $H_5$ and $H_{24}$ respectively.  
The scalars in the {\bf 24} spontaneously break $SU(5)\to SU(3)_C\times SU(2)_L\times U(1)_Y$ while the {\bf 5} contains a Higgs doublet responsible for electroweak symmetry breaking.\footnote{Note that to accomodate neutrino masses, the matter content of the theory must be extended to include an $SU(5)$ singlet, {\em i.e.} the right-handed neutrinos.}  
However, such a minimal setup predicts the light fermion mass ratios to be $m_e/m_\mu\approx m_d/m_s$ which is in contradiction with experimental measurement. 
The solution to this fermion mass ratio problem is to introduce another scalar multiplet, the {\bf 45}~\cite{Georgi:1979df}.  
With this extra multiplet, the correct fermion mass ratio can be achieved.

We take the introduction of scalars in the {\bf 45}, denoted by $H_{45}$, as evidence that there could be addition scalars transforming in some representation of $SU(5)$, denoted by $\Phi$.   Since we want some light components of $\Phi$ to contribute to $A_{FB}^{t\bar{t}}$, $\Phi$ must have a Yukawa coupling to the product of $\Psi_L$ with $\Psi_L$.  Thus $\Phi$ could either be in the 45 or the 50 representation.\footnote{We ignore the possibility that $\Phi$ is in the {\bf 5} representation.}

\subsection{Possible Light Scalar Representations} 
In this subsection we will explore possible light components of $\Phi$ that could unify the SM gauge couplings.  
Recall that the decomposition of the {\bf 45} and {\bf 50} under the SM gauge group is
\begin{equation}
\begin{split}
	45 &= (8,2)_{1/2} \oplus (\bar{6},1)_{-1/3} \oplus (3,3)_{-1/3} \oplus (\bar{3},2)_{-7/6} \oplus (\bar{3},1)_{4/3} \oplus (3,1)_{-1/3} \oplus (1,2)_{1/2},\\
	50 &=  (8,2)_{1/2} \oplus (\bar{6},3)_{-1/3} \oplus (6,1)_{4/3} \oplus (\bar{3},2)_{-7/6} \oplus (3,1)_{-1/3} \oplus (1,1)_{2}.
\end{split}
\end{equation}
To avoid problems with light scalars mediating proton decay, we consider only the light scalars that couple to quarks but not leptons.  
For the {\bf 45}, the qualified components are the $(8,2)_{1/2}$ and $(\bar{6},1)_{-1/3}$, while for the {\bf 50} the qualified components are the $(8,2)_{1/2}$, $(\bar{6},3)_{-1/3}$ and $(6,1)_{4/3}$.  
Now we are ready to address the issue of gauge coupling unification in the presence of these light scalar fields. 

\subsection{Gauge Coupling Evolution}
\label{subsec:gaugecouplingsevolution}
\begin{figure}
	\subfloat[SM]{\label{fig:runningSM}\includegraphics[width=0.5\textwidth]{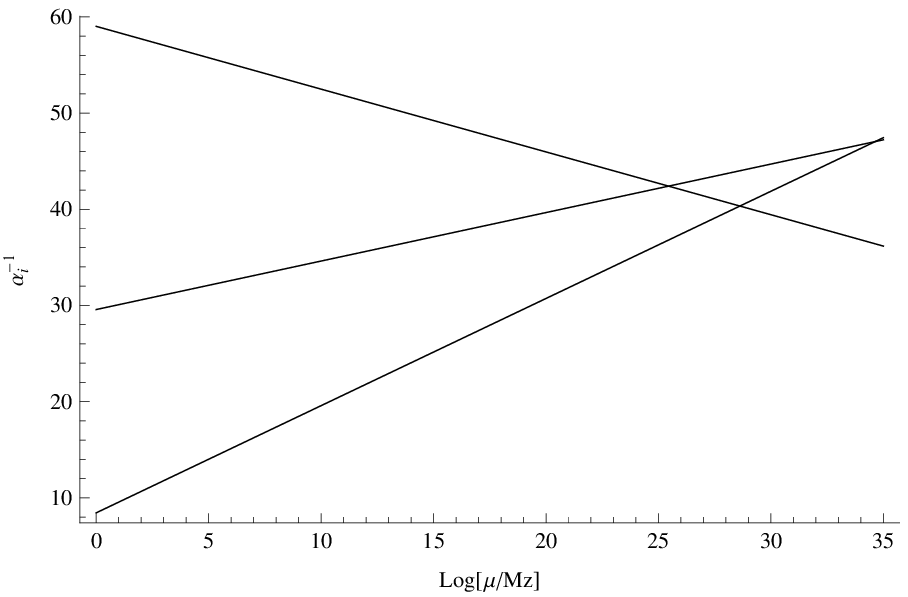}}
	\subfloat[Model with $\Phi$ in the 50]{\label{fig:running50}\includegraphics[width=0.5\textwidth]{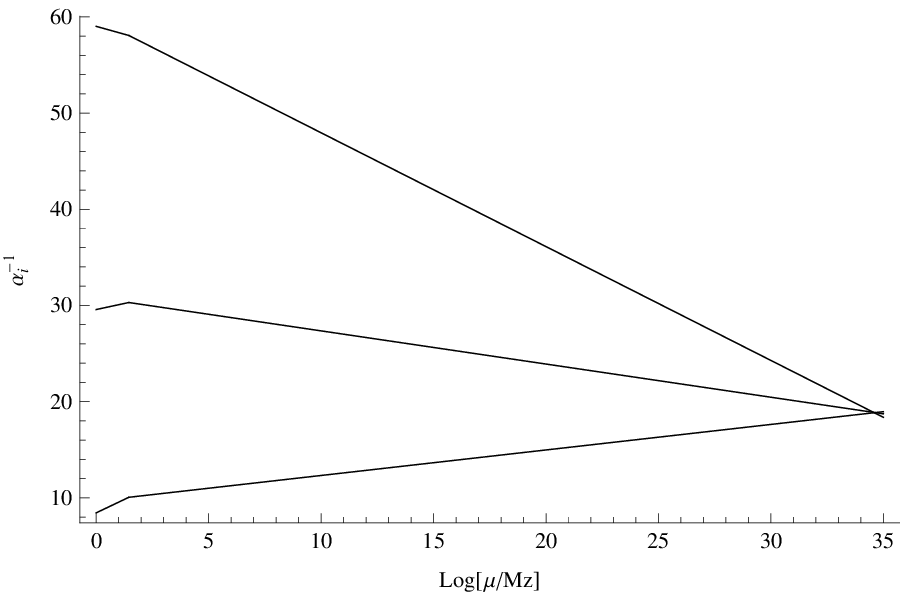}}
	\caption{Gauge couplings running. $\alpha_i=g_i^2/4\pi$}
	\label{fig:running}
\end{figure}

The evolution of gauge couplings is governed by the $\beta$-functions.  At 1-loop level the running of the couplings in the presence of additional scalar particles is given by
\begin{equation}
	\alpha_i^{-1}(t) = \alpha_i^{-1}(M_Z) + \frac{b_i}{2\pi}t + \sum_{t_i}\Theta(t-t_i)\frac{\delta b_i}{2\pi}(t-t_i),
\end{equation}
where $t = \ln(\mu/M_Z)$, $\Theta$ is the Heaviside function, $t_i$ is the scale that new scalars start to contribute and $\delta b_i$ is the contribution due to these new scalars.  The coefficients of the $\beta$ functions from the SM fields are (with 3 generations of fermions and the SU(5) normalization for U$(1)_Y$)
\begin{equation}
\begin{split}
	(b_3, b_2, b_1) &= \left(7,\frac{19}{6},-\frac{41}{10}\right),
\end{split}
\end{equation} 
while the contributions from additional scalar fields are
\begin{equation}
\begin{aligned}
	\left(\delta b_3, \delta b_2, \delta b_1\right)_{(8,2)_{1/2}} &= \left(-2,-\frac{4}{3},-\frac{4}{5}\right),\\
	\left(\delta b_3, \delta b_2, \delta b_1\right)_{(\bar{6},1)_{-1/3}} &= \left(-\frac{5}{6},0,-\frac{2}{15}\right),\\
	\left(\delta b_3, \delta b_2, \delta b_1\right)_{(\bar{6},3)_{-1/3}} &= \left(-\frac{5}{2},-4,-\frac{2}{5}\right),\\
	\left(\delta b_3, \delta b_2, \delta b_1\right)_{(6,1)_{4/3}} &= \left(-\frac{5}{6},0,-\frac{32}{15}\right).
\end{aligned}
\end{equation} 
We found that in the case where $\Phi$ is in the {\bf 45}, the SM with additional light $(8,2)_{1/2}$ and $(\bar{6},1)_{-1/3}$ scalar fields do not lead to gauge coupling unification without having an additional particle at an intermediate scale.\footnote{With an addition of $(3,3)_{-1/3}$ at low scale, one could achieve gauge coupling unifications.  However, the $(3,3)_{-1/3}$ could mediate proton decay unless $\Phi$ is constrained to have Yukawa coupling with $\Psi_L$ but not $\chi_R$.  We will not pursue this possibility in this paper.} 
However, this is not the case for $\Phi$ in the {\bf 50}.  
The SM fields with an additional light $(8,2)_{1/2}$, $(\bar{6},3)_{-1/3}$ and $(6,1)_{4/3}$ lead to gauge coupling unification at the scale $\sim10^{17}$ GeV when the masses of these extra scalar fields are taken to be around 500 GeV, see FIG.~\ref{fig:running}.
Note that this is actually an improvement over typical minimal supersymmetric SM (MSSM) unification scales of $\sim 2\times10^{16}$ GeV~\cite{Dimopoulos:1981yj,Amaldi:1991cn}.
Thus we will ignore the case where $\Phi$ is in the {\bf 45} and focus only on the case where $\Phi$ is in the {\bf 50}.

Note that all we need to achieve gauge coupling unification is to have $(8,2)_{1/2}$, $(\bar{6},3)_{-1/3}$ and $(6,1)_{4/3}$ at a low scale.  
However, they all don't have to come from the same multiplet.  
For example, it is equally valid to have the $(8,2)_{1/2}$ in the same multiplet as the $H_{45}$ while the $(\bar{6},3)_{-1/3}$ and $(6,1)_{4/3}$ are part of $\Phi$ which is in the {\bf 50} representation.  
Since we are interested in having these light scalars mediating a positive $A_{FB}^{t\bar{t}}$, and it is well known in the literature that $(8,2)_{1/2}$ leads to a negative $A_{FB}^{t\bar{t}}$~\cite{Shu:2009xf}, we will focus only on the case where $(8,2)_{1/2}$ is part of the $H_{45}$ while  $(\bar{6},3)_{-1/3}$ and $(6,1)_{4/3}$ are part of the $\Phi$.  
Then we can take the Yukawa coupling of $H_{45}$ to be negligible with impunity. 
After all, such a coupling must be small in order to effect only light fermion mass ratios and not the heavier fermion masses.

Finally we note that the masses of the $(\bar{6},3)_{-1/3}$ and $(6,1)_{4/3}$  can be arranged to be close to the weak scale while the other components remain at GUT scale, see Appendix~\ref{sec:masssplitting} for more details.
\subsection{Yukawa Couplings of Light Scalars} 
To have a consistent GUT model, the Yukawa couplings of light scalars at a low scale cannot be arbitrary.  
In particular, they must remain perturbative and unify at the GUT scale. 
Put another way, the Yukawa couplings at a low scale are determined from the GUT scale Yukawa coupling by the renormalization group (RG) running.  
In this subsection we compute the Yukawa couplings of these light scalars at a low scale via RG running down from the GUT scale.  
The Yukawa coupling at the GUT scale of the {\bf 50} ($\Phi$) and two $\Psi_L$s is
\begin{equation}
	\mathcal{L}_{\Phi} = \frac{Y_G^{ab}}{2}\Psi_{aAB}\Psi_{bCD}\Phi^{AB,CD},
\end{equation}
where $\Phi^{AB,CD} = \Phi^{CD,AB} = -\Phi^{BA,CD} = -\Phi^{AB,DC}$ and $\epsilon_{EABCD}\Phi^{AB,CD}=0$.  
Here $A,B$ denote $SU(5)$ fundamental indices while $a,b$ are flavor indices.  
We denote the light components of $\Phi$ by $\phi_1 = (\bar{6},3)_{-1/3}$, $\phi_2 = (6,1)_{4/3}$ and $\phi_3 = (8,2)_{1/2}$.  
Projecting the Lagrangian onto the basis of light fields yields
\begin{equation}
\label{eq:lagrangian50su2}
\begin{split}
	\mathcal{L}_{\Phi} &= \frac{Y_{\bar{6}}^{ab}}{2} q_{La\alpha}^T Cq_{Lb\beta}\phi_1^{\alpha\beta}+\frac{Y_{6}^{ab}}{2} \overline{u_R}^i_a C\overline{u_R}^j_b\phi_{2ij} +\hc,
\end{split}
\end{equation}
where $C = i\gamma^0\gamma^2$ is the charge conjugation matrix, $i,j$ are $SU(3)_C$ indices and $\alpha, \beta$ are $SU(2)_L$ indices.  
Here the $q_L$ are the left-handed $SU(2)_L$ quark doublets, $q_L = \left( \begin{array}{c} u^i \\ d^i \end{array}\right)$, where the $u^i$ ($d^i$) are the $i^{th}$ generation of the up-type (down-type) quarks.
Similarly, the $u_R$ are the right-handed $SU(2)_L$ singlets.

In general, the Yukawa couplings can be any $3\times3$ symmetric matrices in flavor space.  
However, to avoid problems with flavor changing neutral currents in the light quark sector, we take the Yukawa matrix at the GUT scale to be 
\begin{equation}
	Y_6^{ab} = Y_6\begin{pmatrix}0 & 0 & 1\\ 0 & 0 &0\\ 1 & 0 & 0\end{pmatrix}
	\quad \text{and} \quad
	Y_{\bar{6}}^{ab} = Y_{\bar{6}}\begin{pmatrix}0 & 0 & 1\\ 0 & 0 &0\\ 1 & 0 & 0\end{pmatrix}
\end{equation}  
This structure is preserved by renormalization.  We computed the one-loop running of the Yukawa couplings, assuming that only the top Yukawa coupling, $Y_t$, and  $\Phi$ Yukawa coupling, $Y_i$'s, are sizable.  The relevant $\beta$-functions are
\begin{equation}
\begin{split}
	(2\pi)\frac{\text{d}\alpha_{y_t}}{\text{d}t} &= \left(\frac{9}{2}\alpha_{y_t}+\alpha_{y_6}+\frac{3}{2}\alpha_{y_{\bar{6}}}  - \left(8\alpha_3+\frac{9}{4}\alpha_2+\frac{17}{20}\alpha_1\right)\right)\alpha_{y_t},\\
	(2\pi)\frac{\text{d}\alpha_{y_6}}{\text{d}t} &=\left(4\alpha_{y_6}+2\alpha_{y_t}  - \left(8\alpha_3+\frac{8}{5}\alpha_1\right)\right)\alpha_{y_6},\\
	(2\pi)\frac{\text{d}\alpha_{y_{\bar{6}}}}{\text{d}t} &=\left(5\alpha_{y_{\bar{6}}}+\alpha_{y_t}  - \left(8\alpha_3+\frac{9}{2}\alpha_2+\frac{1}{10}\alpha_1\right)\right)\alpha_{y_{\bar{6}}},
\end{split}
\end{equation}
where we take $\alpha_{y_t} = \frac{Y_t^2}{4\pi}$, $\alpha_{y_6} = \frac{Y_6^2}{4\pi}$ and $\alpha_{y_{\bar{6}}} = \frac{Y_{\bar{6}}^2}{4\pi}$.

Typical values of the new Yukawas at $M_Z$ are computed for various perturbative GUT Yukawas in Table~\ref{yukawas}.
They indicate that reasonable Yukawas at $M_Z$ can yield a large $A_{FB}^{t\bar{t}}$, as we will see in section~\ref{sec:phenom}.
\begin{table}
\begin{tabular}{ | c || c | c | c | c | c | c |}
	\hline
	$Y_G$ & 0.05 & 0.1  & 0.5 & 1 & 1.25& 1.5\\ \hline
	$\alpha_{y_6}$ &0.0049 &0.0181 &0.1173 &0.1433 & 0.1474 &0.1497\\ \hline
	$\alpha_{y_{\bar{6}}}$ & 0.0119 &0.0392 &0.1455 &0.1590 &0.1608  &0.1618\\ \hline
\end{tabular}
\caption{Running of $SU(5)$ Yukawa couplings at GUT scale ($Y_G$) to $M_z$ scale.}
\label{yukawas}
\end{table}

\section{$t\bar{t}$ Phenomenology at the Tevatron}
\label{sec:phenom}
\subsection{General Considerations of $A_{FB}^{t\bar{t}}$}
The forward-backward asymmetry is defined to be
\begin{equation}\label{fbaequation}
	A_{FB}^{t\bar{t}} = \frac{\sigma^{t\bar{t}}_F - \sigma^{t\bar{t}}_B}{\sigma^{t\bar{t}}_{tot}},  
\end{equation}
where forward and backward are defined with respect to the direction of the proton.
In the presence of NP, it is convenient to characterize the asymmetry in terms of the SM and NP contributions.  
We follow~\cite{Grinstein:2011dz} to define $A_{FB}^{t\bar{t}}$ as
\begin{equation}
	A_{FB}^{NP+SM} = \frac{\sigma_F^{NP} - \sigma_B^{NP}}{(\sigma^{NP+SM})_{LO}} + A_{FB}^{SM}\left(\frac{\sigma^{SM}}{\sigma^{SM}+\sigma^{NP}}\right).
\end{equation} 
Note that the first term comes from the leading effect of NP while the second term is the dilution of $A_{FB}^{SM}$ due to NP.  
The observed $A_{FB}^{t\bar{t}}$ reported by the CDF collaboration is $A^{CDF}_{FB} = 0.201\pm0.065_{\text{stat}}\pm0.018_{\text{sys}} = 0.201\pm0.067$~\cite{CDFnote:10584}, where we have combined the uncertainties in quadrature.  
D$\O$ reports a value of $A^{D\O}_{FB} = 0.196\pm0.065$~\cite{Abazov:2011rq}.
The asymmetry predicted by SM is estimated to be $0.073$~\cite{Ahrens:2010zv}, which is about $2\sigma$ away from either observed value.  
However, CDF observed that the asymmetry increased with energy, with $A^{CDF}_{FB} = 0.475\pm0.114$ for $M_{t\bar{t}}\ge 450$ GeV~\cite{Aaltonen:2011kc}.  
The corresponding SM prediction is $0.111$~\cite{Ahrens:2010zv}, a $3.5\sigma$  deviation.  
We take this discrepancy as a hint for NP.

It is worth mentioning that any NP models that wish to explain the $A_{FB}^{t\bar{t}}$ must not violate the measured $t\bar{t}$  production cross-section, $\sigma^{t\bar{t}}$.  
The latest measurement reported by the CDF is $\sigma_{t\bar{t}} = 8.5\pm0.6_{\text{stat}}\pm0.7_{\text{sys}} = 8.5\pm0.9$ pb~\cite{Aaltonen:2010hza}. 
This is to be compared with the SM prediction of $\sigma^{SM} = 6.63$ pb~\cite{Ahrens:2010zv}.  
For further reference, we compile the CDF and D$\O$ measurements as well as the SM prediction~\cite{Ahrens:2010zv} in Table~\ref{tab:observables}.
\begin{table}
\begin{tabular}{| c || c | c |}
	\hline
	Observable & Measured Value & SM Prediction~\cite{Ahrens:2010zv} \\ \hline \hline
	$A_{FB}^{t\bar{t}}$ & \begin{tabular}{c}$0.196\pm0.065$~\cite{Abazov:2011rq} \\ $0.201\pm0.065_{\text{stat}}\pm0.018_{\text{sys}}$~\cite{CDFnote:10584} \end{tabular}& $0.073$ \\ \hline
	$A_{FB}^{t\bar{t}}(M_{t\bar{t}}\le 450$ GeV) & $-0.116\pm0.153$~\cite{Aaltonen:2011kc} & $0.052$ \\ \hline
	$A_{FB}^{t\bar{t}}(M_{t\bar{t}}\ge 450$ GeV) & $0.475\pm0.114$~\cite{Aaltonen:2011kc} & $0.111$ \\ \hline
	$\sigma^{t\bar{t}}$ & $8.5\pm0.6_{\text{stat}}\pm0.7_{\text{sys}}$ pb~\cite{Aaltonen:2010hza} &  $6.63$ \\ \hline
\end{tabular}
\caption{Measurements and SM predictions of $t\bar{t}$ observables at the Tevatron.}
\label{tab:observables}
\end{table}


\subsection{Differential Cross-section for $t\bar{t}$ Production}
To study $t\bar{t}$ phenomenology, it is convenient to expand the $SU(2)$ indices in the above Lagrangian, Eq. (\ref{eq:lagrangian50su2}), and keep terms relevant for the tree-level $t\bar{t}$ production cross-section:
\begin{equation}
\begin{split}
	\mathcal{L}_{\Phi} &= \frac{Y_{\bar{6}}^{ut}+Y_{\bar{6}}^{tu}}{2}(u^TCP_Lt\phi_1^1+\frac{1}{\sqrt{2}}d^TCP_Lt\phi_1^2) + \frac{Y_6^{ut}+Y_6^{tu}}{2}\bar{u}CP_L\bar{t}^T\phi_2+\hc,
\end{split}
\end{equation} 
where in the above expression $SU(3)_C$ indices have been suppressed. We take the $SU(2)$ two-index symmetric tensor to be $\phi_1 = \begin{pmatrix}\phi_1^1 &\phi_1^2/\sqrt{2} \\ \phi_1^2/\sqrt{2} &\phi_1^3\end{pmatrix}$.  
At the Tevatron, $t\bar{t}$ are dominantly produced from the $u$ quarks or the $d$ quarks, while other quarks or gluon initial states are PDF suppressed. The differential cross-section for $t\bar{t}$ production initiated from the $u$ quarks is
\begin{align}
	\frac{\text{d}\sigma^{(NP)}_{(\bar{6},3)}}{\text{d}\hat{t}}(u\bar{u}\to t\bar{t}) &= \frac{1}{16\pi\hat{s}^2}\frac{1}{4}\frac{1}{9}\left(-\frac{8g_s^2(4\pi\alpha_{y_{\bar{6}}})}{s(\hat{u}-m_{\phi_1}^2)}\left((\hat{u}-m_t^2)^2+\hat{s}m_t^2\right)\right.\nn\\
	&\hspace{2.5cm} \left. + 4(4\pi\alpha_{y_{\bar{6}}})^2\frac{3}{2}\frac{(\hat{u}-m_t^2)^2}{(\hat{u}-m^2_{\phi_1})^2}\right), \displaybreak[0]\\ 
	\frac{\text{d}\sigma^{(NP)}_{(6,1)}}{\text{d}\hat{t}}(u\bar{u}\to t\bar{t}) &= \frac{1}{16\pi\hat{s}^2}\frac{1}{4}\frac{1}{9}\left(-\frac{8g_s^2(4\pi\alpha_{y_6})}{s(\hat{u}-m_{\phi_2}^2)}\left((\hat{u}-m_t^2)^2+\hat{s}m_t^2\right)\right.\nn\\
	&\hspace{2.5cm} \left. + 4(4\pi\alpha_{y_6})^2\frac{3}{2}\frac{(\hat{u}-m_t^2)^2}{(\hat{u}-m^2_{\phi_2})^2}\right), 
\end{align}
where we have defined $\alpha_{y_{\bar{6}}} =  \left(\frac{Y_{\bar{6}}^{ut}+Y_{\bar{6}}^{tu}}{2}\right)^2/4\pi$, and  $\alpha_{y_6}$ is also defined analogously.  Note that we have included interference with the SM in our NP cross-section.  Similarly, the differential cross-section initiated from the $d$ quarks is
\begin{align}
	\frac{\text{d}\sigma^{(NP)}_{(\bar{6},3)}}{\text{d}\hat{t}}(d\bar{d}\to t\bar{t}) &= \frac{1}{16\pi\hat{s}^2}\frac{1}{4}\frac{1}{9}\left(-\frac{4g_s^2(4\pi\alpha_{y_{\bar{6}}})}{s(\hat{u}-m_{\phi_1}^2)}\left((\hat{u}-m_t^2)^2+\hat{s}m_t^2\right)\right.\nn\\
	&\hspace{2.5cm} \left. + (4\pi\alpha_{y_{\bar{6}}})^2\frac{3}{2}\frac{(\hat{u}-m_t^2)^2}{(\hat{u}-m^2_{\phi_1})^2}\right).
\end{align}


\subsection{Numerical Results from the Model}
The above differential cross-sections must be convoluted with parton distribution functions (PDFs) of the proton and anti-proton to give $A_{FB}^{t\bar{t}}$ comparable with accelerator measurements.

\begin{center}
\begin{figure}
	\begin{center}
	\subfloat[High invariant mass bin asymmetry]{\label{fig:yg05afbhi}\psfrag{A}{}\psfrag{B}{$A^{t\bar{t}}_{FB}(M_{t\bar{t}} > 450 \text{GeV})$}\psfrag{xlabelpsf}{$m_6$ [GeV]}\includegraphics[width=0.45\textwidth]{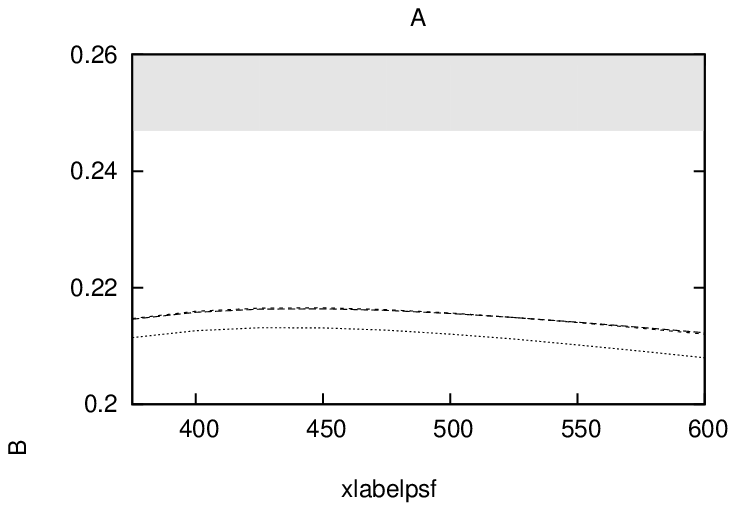}}
	\hfill
	\subfloat[Low invariant mass bin asymmetry]{\label{fig:yg05afblow}\psfrag{A}{}\psfrag{B}{$A^{t\bar{t}}_{FB}(M_{t\bar{t}} < 450 \text{GeV})$}\psfrag{xlabelpsf}{$m_6$ [GeV]}\includegraphics[width=0.45\textwidth]{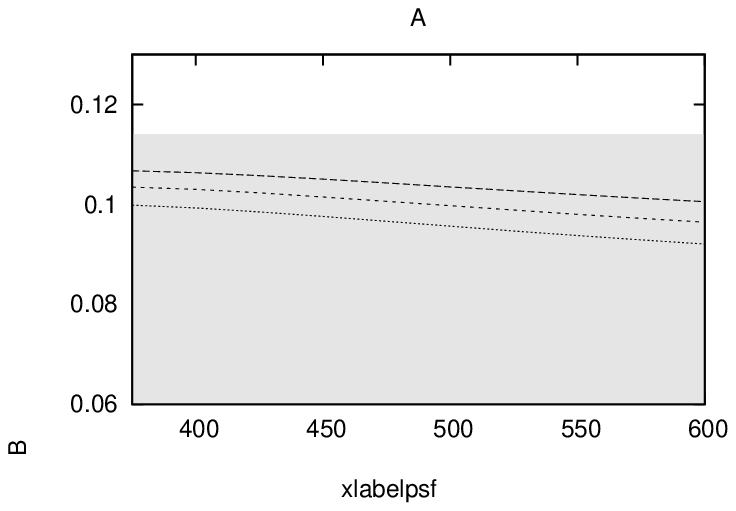}}\\
	\subfloat[Total asymmetry]{\label{fig:yg05afbtot}\psfrag{A}{}\psfrag{B}{$A^{t\bar{t}}_{FB}$}\psfrag{xlabelpsf}{$m_6$[GeV]}\includegraphics[width=0.45\textwidth]{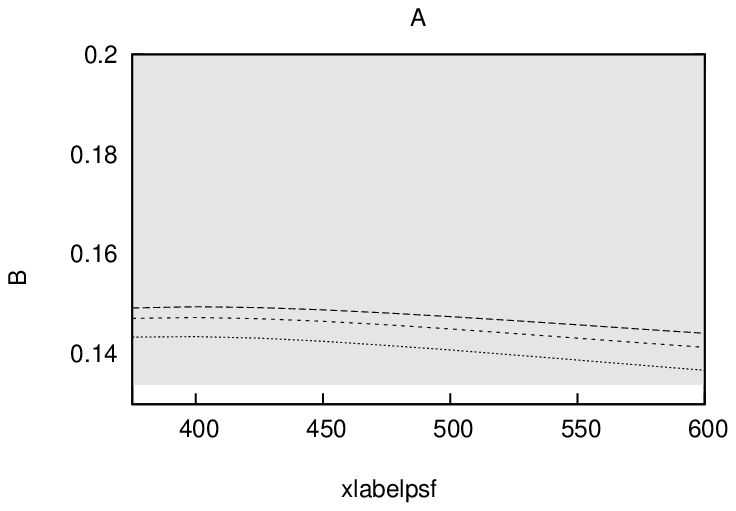}}
	\hfill
	\subfloat[Total (NP+SM) $t\bar{t}$ production]{\label{fig:yg05sigtot}\psfrag{A}{}\psfrag{B}{$\sigma^{NP}_{tot} + \sigma^{SM}_{tot}$[pb]}\psfrag{xlabelpsf}{$m_6$ [GeV]}\includegraphics[width=0.45\textwidth]{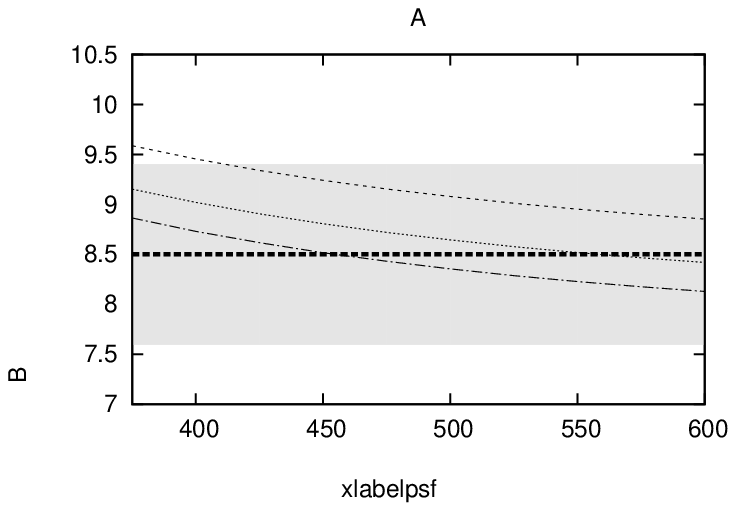}}
	\end{center}
	\caption{Computational results for $t\bar{t}$ phenomenology at a GUT Yukawa coupling of $Y_G = 0.5$. The contours in each plot, from bottom to top, are decreasing in $m_{(\bar{6},3)}$ from 550 to 400 GeV. The gray regions are, in plot~(\ref{fig:yg05afbhi}), CDF 2$\sigma$ allowed regions, in plot~(\ref{fig:yg05afblow}), CDF 1.5$\sigma$ allowed regions, and, in plots~(\ref{fig:yg05afbtot}) and~(\ref{fig:yg05sigtot}), CDF 1$\sigma$ allowed regions. Plot~(\ref{fig:yg05sigtot}) also shows the central value for the CDF $t\bar t$ cross-section. }
	\label{fig:yg05fig}
\end{figure}
\end{center}

\begin{center}
\begin{figure}
	\begin{center}
	\subfloat[High invariant mass bin asymmetry]{\label{fig:yg10afbhi}\psfrag{A}{}\psfrag{B}{$A^{t\bar{t}}_{FB}(M_{t\bar{t}} > 450 \text{ GeV})$}\psfrag{xlabelpsf}{$m_6$ [GeV]}\includegraphics[width=0.45\textwidth]{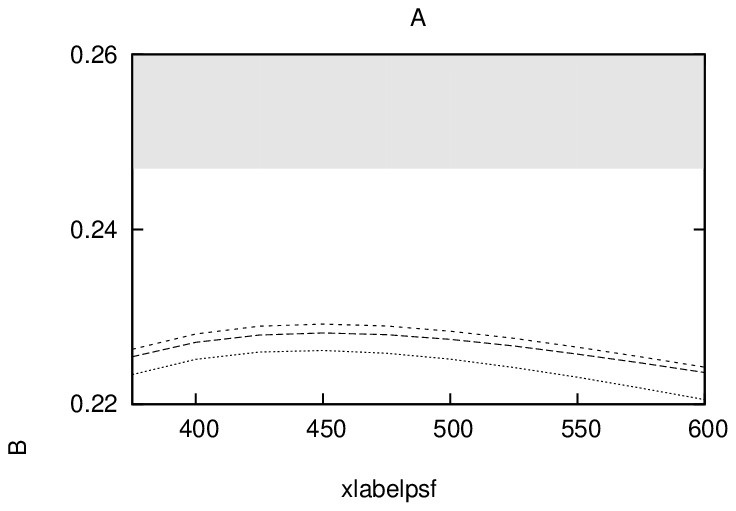}}
	\hfill
	\subfloat[Low invariant mass bin asymmetry]{\label{fig:yg10afblow}\psfrag{A}{}\psfrag{B}{$A^{t\bar{t}}_{FB}(M_{t\bar{t}} < 450 \text{ GeV})$}\psfrag{xlabelpsf}{$m_6$ [GeV]}\includegraphics[width=0.45\textwidth]{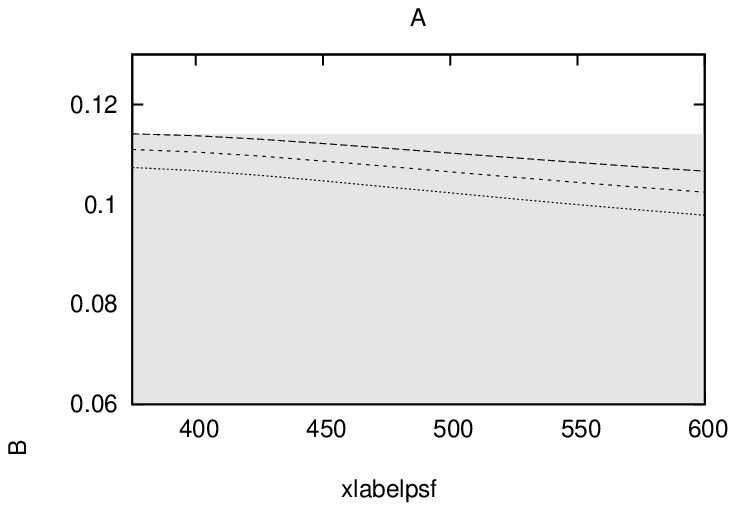}}\\
	\subfloat[Total asymmetry]{\label{fig:yg10afbtot}\psfrag{A}{}\psfrag{B}{$A^{t\bar{t}}_{FB}$}\psfrag{xlabelpsf}{$m_6$ [GeV]}\includegraphics[width=0.45\textwidth]{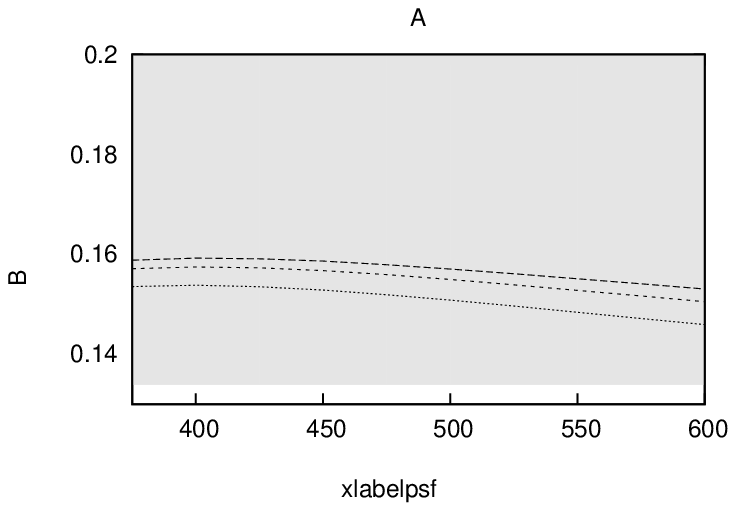}}
	\hfill
	\subfloat[Total (NP+SM) $t\bar{t}$ production]{\label{fig:yg10sigtot}\psfrag{A}{}\psfrag{B}{$\sigma^{NP}_{tot} + \sigma^{SM}_{tot}$ [pb]}\psfrag{xlabelpsf}{$m_6$ [GeV]}\includegraphics[width=0.45\textwidth]{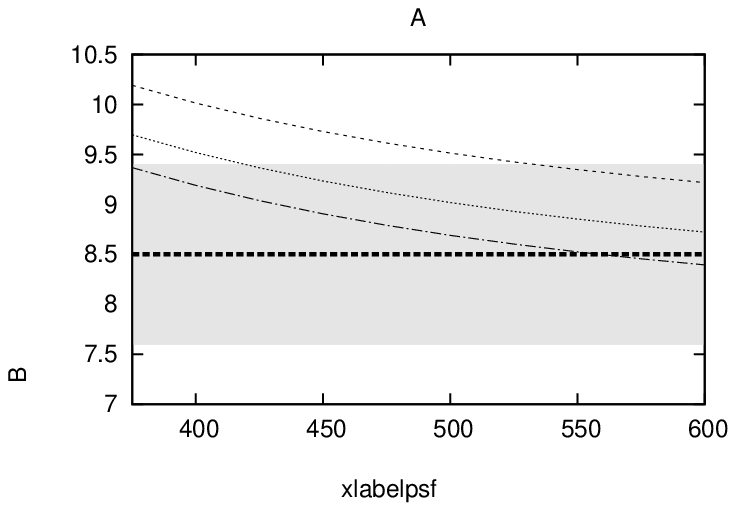}}
	\caption{Same as FIG.~\ref{fig:yg05fig}, except with $Y_G = 1.0$. This coupling provides better agreement with measurements.}
	\label{fig:yg10fig}
	\end{center}
\end{figure}
\end{center}

We compute the total cross-sections and asymmetries using the NLO MSTW 2008 PDFs~\cite{Martin:2009iq}.  We find that for a suitable set of parameters, our model can accommodate the large $A_{FB}^{t\bar{t}}$ and be consistent with the $t\bar{t}$ production cross-section constraint as can be seen in FIGS.~\ref{fig:yg05fig} and~\ref{fig:yg10fig}.

The phenomenological aspects of the model provide nice improvements over SM predictions.
The $\fba$ from our model agrees with CDF data within $\sim2.2\sigma$ for the high-mass bin, $\sim1.4\sigma$ for the low-mass bin, and $\sim0.7\sigma$ for the total asymmetry.
Additionally, our model agrees with the CDF total $t\bar{t}$ production within $\sim0.9\sigma$.
While the high-mass bin result seems most problematic with the model, it is actually the most significant improvement over SM results.
The CDF low-mass bin and total asymmetry measurements are discrepant from the NNLO SM predictions~\cite{Ahrens:2010zv} at $\sim2\sigma$ or less and are not statistically significant.
However, the high-mass bin is discrepant at $\sim3.5\sigma$, and so our model reduces this deviation to a less statistically signficant result.
The agreement within less than $1.4\sigma$ of the rest of the results serves as a check on the merits of the model.
  
\section{Conclusion and Discussion}
\label{sec:conclusions}

Light colored scalar extensions of SM can account for the observed $A_{FB}^{t\bar{t}}$ reported by CDF and D$\O$. 
At the same time they also allow for unification of fundamental forces at sufficiently high scales consistent with bounds from proton decay. 
In our explicit model with a $SU(5)$ GUT, we extend SM by introducing 3 multiplets of light scalars:  $(8,2)_{1/2}\in{\bf 45}$, $(\bar{6},3)_{-1/3}$ and $(6,1)_{4/3}\in{\bf 50}$.  
These new scalars lead to gauge coupling unification at a scale of $10^{17}$ GeV, which is considerably higher compared to typical unification scales suggested by MSSM, $2\times10^{16}$ GeV.  
Notice that the quantum number of these scalars forbids Yukawa coupling to leptons. 
Hence there is no light scalar leptoquark mediating proton decay.  
In this work we do not attempt to solve the hierarchy problem associated with these light scalars.
 
For suitable values of the Yukawa coupling and masses these colored scalars that contribute to $A^{t\bar{t}}_{FB}$ yield values of 9.5-11.5\% (CDF: -11.6$\pm$15.3\%) for $M_{t\bar{t}}\le450$ GeV, 21-22.5\% (CDF: 47.5$\pm$11.4\%) for $M_{t\bar{t}}>450$, and 14-16\% (CDF: 20.7$\pm$6.7\%, D$\O$: $19.6\pm6.5\%$) for the total forward-backward asymmetry.  
The corresponding total $t\bar{t}$ production cross-section, including these scalars' contribution, is around $8.4-10.2$ pb (CDF: $8.5\pm0.9$ pb).
These computations have been checked under variations of the SM input parameters ($m_t$, $\alpha_s$, and PDF parameters) within their reported 1$\sigma$ limits; our results show sub-percent variations and thus the ranges of the values reported above can be trusted. 

It is interesting to explore the LHC phenomenology associated with this model.  
In principle, a single color sextet scalar $\phi\in\{(\bar{6},3)_{-1/3}$ and $(6,1)_{4/3}\}$ can be singly produced from a pair of quark initial states $qq\to\phi$. The single production channel does depends on the form of Yukawa coupling matrix.  In the case where the Yukawa coupling is diagonal and of $\mathcal{O}(1)$, the production cross-section from the $uu$ initial state, $gg \to \phi$, is $\sim 10$ nb~\cite{Han:2009ya}. 
However in our model, due to the particular form of the Yukawa coupling, the possible initial states are $ut$ or $db$.  Thus single $\phi$ production will be suppressed by the $t(b)$ PDF.  After being produced, $\phi$ would decay into a pair of $ut(db)$ which lead to 2 jets with or without leptons.

Nevertheless, the LHC is known as a gluon factory, and thus a more promising production mechanism is pair production from gluon fusion, $gg \to \phi\bar{\phi}$.  The production cross-section in this channel, given the mass of $\phi\approx500$ GeV, is at the order of a few pb~\cite{Chen:2008hh}.
This production channel would lead to 4 jets, 2 of which are $b$ jets, with or without leptons.  
Thus in this case it is possible to observe two widely separated b jets and the invariant mass of these 4 jets displays a resonant structure at $2m_\phi$.

\begin{acknowledgements}
  We thank B. Grinstein for suggesting this project and for various insightful discussions. This work was supported in part by the  US Department of Energy under contract DOE-FG03-97ER40546.
\end{acknowledgements}


\appendix
\section{Mass Splitting in the {\bf 50}}
\label{sec:masssplitting}
In this appendix we show that the $(6,1)_{4/3}$ and $(\bar{6},1)_{-1/3}$ of the {\bf 50} ($\Phi$) can be made arbitrarily light while the masses of the other components remain close to the GUT scale. 
The splitting happens during the spontaneous breaking of $SU(5)\to SU(3)_C\times SU(2)_L\times U(1)_Y$ by the vev of the adjoint scalar, $\vev{H_{24}}=v_{24}\text{diag}(-2,-2,-2,3,3)$.  
For convenience, we label each component of the {\bf 50} by $(\phi_1,\phi_2,\phi_3,\phi_4,\phi_5,\phi_6)$ = ($(\bar{6},3)_{-1/3}$, $(6,1)_{4/3}$, $(8,2)_{1/2}$, $(\bar{3},2)_{-7/6}$, $(3,1)_{-1/3}$, $(1,1)_{2}$).  
To see that the multiplet does indeed split, consider the renormalizable scalar potential of the form
\begin{equation}
\begin{split}
	V(\Phi) &= m_1^2\Phi^{ABCD}\Phi^\dagger_{ABCD} + \frac{m_2^2}{v_{24}}\Phi^{ABCD}\Phi^\dagger_{ABCE}(H_{24})^E_D\\
	&\qquad + \frac{m_3^2}{v_{24}^2}\Phi^{ABCD}\Phi^\dagger_{ABEF}(H_{24})^E_C(H_{24})^F_D,
\end{split}
\end{equation}
where $m_i$s are at their natural value around the GUT scale, $\Phi^{ABCD}$ is antisymmetric in $A\leftrightarrow B$, $C \leftrightarrow D$, symmetric in $(AB)\leftrightarrow(CD)$ with $\epsilon_{EABCD}\Phi^{ABCD} = 0$ and $A,\ldots,F$ are $SU(5)$ indices.  
Expanding around $H_{24}=\vev{H_{24}}$, the mass of each component of $\Phi$ is 
\begin{equation}
\begin{aligned}
	m_{\phi_1}^2 &= m_1^2 + \frac{1}{2}m_2^2 - 6m_3^2,\\
	m_{\phi_2}^2 &= m_1^2 - m_2^2 + 4m_3^2,\\
	m_{\phi_3}^2 &= m_1^2 - \frac{3}{4}m_2^2 - m_3^2,
\end{aligned}
\qquad
\begin{aligned}
	m_{\phi_4}^2 &= m_1^2 + \frac{7}{4}m_2^2 +\frac{3}{2} 6m_3^2,\\
	m_{\phi_5}^2 &= m_1^2 + \frac{1}{2}m_2^2 +\frac{1}{4}m_3^2,\\
	m_{\phi_6}^2 &= m_1^2 + 3m_2^2 + 9m_3^2.
\end{aligned}
\end{equation}
Thus, by tuning $m_1$, $m_2$ and $m_3$, the masses of $(\phi_1,\phi_2)$ = ($(\bar{6},3)_{-1/3}$, $(6,1)_{4/3}$) can be made arbitrarily light while the other components remain heavy.

\bibliography{ttbar}{}

\begin{thebibliography}{22}%
\makeatletter
\providecommand \@ifxundefined [1]{%
 \@ifx{#1\undefined}
}%
\providecommand \@ifnum [1]{%
 \ifnum #1\expandafter \@firstoftwo
 \else \expandafter \@secondoftwo
 \fi
}%
\providecommand \@ifx [1]{%
 \ifx #1\expandafter \@firstoftwo
 \else \expandafter \@secondoftwo
 \fi
}%
\providecommand \natexlab [1]{#1}%
\providecommand \enquote  [1]{``#1''}%
\providecommand \bibnamefont  [1]{#1}%
\providecommand \bibfnamefont [1]{#1}%
\providecommand \citenamefont [1]{#1}%
\providecommand \href@noop [0]{\@secondoftwo}%
\providecommand \href [0]{\begingroup \@sanitize@url \@href}%
\providecommand \@href[1]{\@@startlink{#1}\@@href}%
\providecommand \@@href[1]{\endgroup#1\@@endlink}%
\providecommand \@sanitize@url [0]{\catcode `\\12\catcode `\$12\catcode
  `\&12\catcode `\#12\catcode `\^12\catcode `\_12\catcode `\%12\relax}%
\providecommand \@@startlink[1]{}%
\providecommand \@@endlink[0]{}%
\providecommand \url  [0]{\begingroup\@sanitize@url \@url }%
\providecommand \@url [1]{\endgroup\@href {#1}{\urlprefix }}%
\providecommand \urlprefix  [0]{URL }%
\providecommand \Eprint [0]{\href }%
\@ifxundefined \urlstyle {%
  \providecommand \doi  [0]{\begingroup \@sanitize@url \@doi}%
  \providecommand \@doi [1]{\endgroup \@@startlink {\doibase
  #1}doi:\discretionary {}{}{}#1\@@endlink }%
}{%
  \providecommand \doi  [0]{doi:\discretionary{}{}{}\begingroup
  \urlstyle{rm}\Url }%
}%
\providecommand \doibase [0]{http://dx.doi.org/}%
\providecommand \Doi [0]{\begingroup \@sanitize@url \@Doi }%
\providecommand \@Doi  [1]{\endgroup\@@startlink{\doibase#1}\@@Doi}%
\providecommand \@@Doi [1]{#1\@@endlink}%
\providecommand \selectlanguage [0]{\@gobble}%
\providecommand \bibinfo  [0]{\@secondoftwo}%
\providecommand \bibfield  [0]{\@secondoftwo}%
\providecommand \translation [1]{[#1]}%
\providecommand \BibitemOpen [0]{}%
\providecommand \bibitemStop [0]{}%
\providecommand \bibitemNoStop [0]{.\EOS\space}%
\providecommand \EOS [0]{\spacefactor3000\relax}%
\providecommand \BibitemShut  [1]{\csname bibitem#1\endcsname}%
\bibitem [{\citenamefont {Georgi}\ and\ \citenamefont
  {Glashow}(1974)}]{Georgi:1974sy}%
  \BibitemOpen
  \bibfield  {author} {\bibinfo {author} {\bibfnamefont {H.}~\bibnamefont
  {Georgi}}\ and\ \bibinfo {author} {\bibfnamefont {S.}~\bibnamefont
  {Glashow}},\ }\Doi {10.1103/PhysRevLett.32.438} {\bibfield  {journal}
  {\bibinfo  {journal} {Phys.Rev.Lett.},\ }\textbf {\bibinfo {volume} {32}},\
  \bibinfo {pages} {438} (\bibinfo {year} {1974})}\BibitemShut {NoStop}%
\bibitem [{\citenamefont {Georgi}\ and\ \citenamefont
  {Jarlskog}(1979)}]{Georgi:1979df}%
  \BibitemOpen
  \bibfield  {author} {\bibinfo {author} {\bibfnamefont {H.}~\bibnamefont
  {Georgi}}\ and\ \bibinfo {author} {\bibfnamefont {C.}~\bibnamefont
  {Jarlskog}},\ }\Doi {10.1016/0370-2693(79)90842-6} {\bibfield  {journal}
  {\bibinfo  {journal} {Phys.Lett.},\ }\textbf {\bibinfo {volume} {B86}},\
  \bibinfo {pages} {297} (\bibinfo {year} {1979})}\BibitemShut {NoStop}%
\bibitem [{\citenamefont {Abazov}\ \emph {et~al.}(2011)\citenamefont {Abazov}
  \emph {et~al.}}]{Abazov:2011rq}%
  \BibitemOpen
  \bibfield  {author} {\bibinfo {author} {\bibfnamefont {V.~M.}\ \bibnamefont
  {Abazov}} \emph {et~al.} (\bibinfo {collaboration} {D0 Collaboration}),\
  }\Doi {10.1103/PhysRevD.84.112005} {\bibfield  {journal} {\bibinfo  {journal}
  {Phys.Rev.},\ }\textbf {\bibinfo {volume} {D84}},\ \bibinfo {pages} {112005}
  (\bibinfo {year} {2011})},\ \Eprint {http://arxiv.org/abs/1107.4995}
  {arXiv:1107.4995 [hep-ex]} \BibitemShut {NoStop}%
\bibitem [{\citenamefont {Aaltonen}\ \emph
  {et~al.}(2011){\natexlab{a}}\citenamefont {Aaltonen} \emph
  {et~al.}}]{Aaltonen:2011kc}%
  \BibitemOpen
  \bibfield  {author} {\bibinfo {author} {\bibfnamefont {T.}~\bibnamefont
  {Aaltonen}} \emph {et~al.} (\bibinfo {collaboration} {CDF Collaboration}),\
  }\Doi {10.1103/PhysRevD.83.112003} {\bibfield  {journal} {\bibinfo  {journal}
  {Phys.Rev.},\ }\textbf {\bibinfo {volume} {D83}},\ \bibinfo {pages} {112003}
  (\bibinfo {year} {2011}{\natexlab{a}})},\ \Eprint
  {http://arxiv.org/abs/1101.0034} {arXiv:1101.0034 [hep-ex]} \BibitemShut
  {NoStop}%
\bibitem [{\citenamefont {Schwarz}\ \emph {et~al.}(2011)\citenamefont {Schwarz}
  \emph {et~al.}}]{CDFnote:10584}%
  \BibitemOpen
  \bibfield  {author} {\bibinfo {author} {\bibfnamefont {T.}~\bibnamefont
  {Schwarz}} \emph {et~al.} (\bibinfo {collaboration} {CDF Collaboration}),\
  }\href@noop {} { (\bibinfo {year} {2011})},\ \Eprint
  {http://arxiv.org/abs/CDF Note 10584} {CDF Note 10584} \BibitemShut {NoStop}%
\bibitem [{\citenamefont {Ahrens}\ \emph {et~al.}(2010)\citenamefont {Ahrens},
  \citenamefont {Ferroglia}, \citenamefont {Neubert}, \citenamefont {Pecjak},\
  and\ \citenamefont {Yang}}]{Ahrens:2010zv}%
  \BibitemOpen
  \bibfield  {author} {\bibinfo {author} {\bibfnamefont {V.}~\bibnamefont
  {Ahrens}}, \bibinfo {author} {\bibfnamefont {A.}~\bibnamefont {Ferroglia}},
  \bibinfo {author} {\bibfnamefont {M.}~\bibnamefont {Neubert}}, \bibinfo
  {author} {\bibfnamefont {B.~D.}\ \bibnamefont {Pecjak}}, \ and\ \bibinfo
  {author} {\bibfnamefont {L.~L.}\ \bibnamefont {Yang}},\ }\Doi
  {10.1007/JHEP09(2010)097} {\bibfield  {journal} {\bibinfo  {journal} {JHEP},\
  }\textbf {\bibinfo {volume} {1009}},\ \bibinfo {pages} {097} (\bibinfo {year}
  {2010})},\ \Eprint {http://arxiv.org/abs/1003.5827} {arXiv:1003.5827
  [hep-ph]} \BibitemShut {NoStop}%
\bibitem [{\citenamefont {Shu}\ \emph {et~al.}(2010)\citenamefont {Shu},
  \citenamefont {Tait},\ and\ \citenamefont {Wang}}]{Shu:2009xf}%
  \BibitemOpen
  \bibfield  {author} {\bibinfo {author} {\bibfnamefont {J.}~\bibnamefont
  {Shu}}, \bibinfo {author} {\bibfnamefont {T.~M.}\ \bibnamefont {Tait}}, \
  and\ \bibinfo {author} {\bibfnamefont {K.}~\bibnamefont {Wang}},\ }\Doi
  {10.1103/PhysRevD.81.034012} {\bibfield  {journal} {\bibinfo  {journal}
  {Phys.Rev.},\ }\textbf {\bibinfo {volume} {D81}},\ \bibinfo {pages} {034012}
  (\bibinfo {year} {2010})},\ \Eprint {http://arxiv.org/abs/0911.3237}
  {arXiv:0911.3237 [hep-ph]} \BibitemShut {NoStop}%
\bibitem [{\citenamefont {Dorsner}\ \emph {et~al.}(2010)\citenamefont
  {Dorsner}, \citenamefont {Fajfer}, \citenamefont {Kamenik},\ and\
  \citenamefont {Kosnik}}]{Dorsner:2009mq}%
  \BibitemOpen
  \bibfield  {author} {\bibinfo {author} {\bibfnamefont {I.}~\bibnamefont
  {Dorsner}}, \bibinfo {author} {\bibfnamefont {S.}~\bibnamefont {Fajfer}},
  \bibinfo {author} {\bibfnamefont {J.~F.}\ \bibnamefont {Kamenik}}, \ and\
  \bibinfo {author} {\bibfnamefont {N.}~\bibnamefont {Kosnik}},\ }\Doi
  {10.1103/PhysRevD.81.055009} {\bibfield  {journal} {\bibinfo  {journal}
  {Phys.Rev.},\ }\textbf {\bibinfo {volume} {D81}},\ \bibinfo {pages} {055009}
  (\bibinfo {year} {2010})},\ \Eprint {http://arxiv.org/abs/0912.0972}
  {arXiv:0912.0972 [hep-ph]} \BibitemShut {NoStop}%
\bibitem [{\citenamefont {Barger}\ \emph {et~al.}(2010)\citenamefont {Barger},
  \citenamefont {Keung},\ and\ \citenamefont {Yu}}]{Barger:2010mw}%
  \BibitemOpen
  \bibfield  {author} {\bibinfo {author} {\bibfnamefont {V.}~\bibnamefont
  {Barger}}, \bibinfo {author} {\bibfnamefont {W.-Y.}\ \bibnamefont {Keung}}, \
  and\ \bibinfo {author} {\bibfnamefont {C.-T.}\ \bibnamefont {Yu}},\ }\Doi
  {10.1103/PhysRevD.81.113009} {\bibfield  {journal} {\bibinfo  {journal}
  {Phys.Rev.},\ }\textbf {\bibinfo {volume} {D81}},\ \bibinfo {pages} {113009}
  (\bibinfo {year} {2010})},\ \Eprint {http://arxiv.org/abs/1002.1048}
  {arXiv:1002.1048 [hep-ph]} \BibitemShut {NoStop}%
\bibitem [{\citenamefont {Dorsner}\ \emph {et~al.}(2011)\citenamefont
  {Dorsner}, \citenamefont {Drobnak}, \citenamefont {Fajfer}, \citenamefont
  {Kamenik},\ and\ \citenamefont {Kosnik}}]{Dorsner:2011ai}%
  \BibitemOpen
  \bibfield  {author} {\bibinfo {author} {\bibfnamefont {I.}~\bibnamefont
  {Dorsner}}, \bibinfo {author} {\bibfnamefont {J.}~\bibnamefont {Drobnak}},
  \bibinfo {author} {\bibfnamefont {S.}~\bibnamefont {Fajfer}}, \bibinfo
  {author} {\bibfnamefont {J.~F.}\ \bibnamefont {Kamenik}}, \ and\ \bibinfo
  {author} {\bibfnamefont {N.}~\bibnamefont {Kosnik}},\ }\Doi
  {10.1007/JHEP11(2011)002} {\bibfield  {journal} {\bibinfo  {journal} {JHEP},\
  }\textbf {\bibinfo {volume} {1111}},\ \bibinfo {pages} {002} (\bibinfo {year}
  {2011})},\ \Eprint {http://arxiv.org/abs/1107.5393} {arXiv:1107.5393
  [hep-ph]} \BibitemShut {NoStop}%
\bibitem [{\citenamefont {Shelton}\ and\ \citenamefont
  {Zurek}(2011)}]{Shelton:2011hq}%
  \BibitemOpen
  \bibfield  {author} {\bibinfo {author} {\bibfnamefont {J.}~\bibnamefont
  {Shelton}}\ and\ \bibinfo {author} {\bibfnamefont {K.~M.}\ \bibnamefont
  {Zurek}},\ }\Doi {10.1103/PhysRevD.83.091701} {\bibfield  {journal} {\bibinfo
   {journal} {Phys.Rev.},\ }\textbf {\bibinfo {volume} {D83}},\ \bibinfo
  {pages} {091701} (\bibinfo {year} {2011})},\ \Eprint
  {http://arxiv.org/abs/1101.5392} {arXiv:1101.5392 [hep-ph]} \BibitemShut
  {NoStop}%
\bibitem [{\citenamefont {Nelson}\ \emph {et~al.}(2011)\citenamefont {Nelson},
  \citenamefont {Okui},\ and\ \citenamefont {Roy}}]{Nelson:2011us}%
  \BibitemOpen
  \bibfield  {author} {\bibinfo {author} {\bibfnamefont {A.~E.}\ \bibnamefont
  {Nelson}}, \bibinfo {author} {\bibfnamefont {T.}~\bibnamefont {Okui}}, \ and\
  \bibinfo {author} {\bibfnamefont {T.~S.}\ \bibnamefont {Roy}},\ }\href@noop
  {} {\bibfield  {journal} {\bibinfo  {journal} {Phys.Rev.},\ }\textbf
  {\bibinfo {volume} {D84}},\ \bibinfo {pages} {094007} (\bibinfo {year}
  {2011})},\ \Eprint {http://arxiv.org/abs/1104.2030} {arXiv:1104.2030
  [hep-ph]} \BibitemShut {NoStop}%
\bibitem [{\citenamefont {Grinstein}\ \emph
  {et~al.}(2011){\natexlab{a}}\citenamefont {Grinstein}, \citenamefont {Kagan},
  \citenamefont {Trott},\ and\ \citenamefont {Zupan}}]{Grinstein:2011yv}%
  \BibitemOpen
  \bibfield  {author} {\bibinfo {author} {\bibfnamefont {B.}~\bibnamefont
  {Grinstein}}, \bibinfo {author} {\bibfnamefont {A.~L.}\ \bibnamefont
  {Kagan}}, \bibinfo {author} {\bibfnamefont {M.}~\bibnamefont {Trott}}, \ and\
  \bibinfo {author} {\bibfnamefont {J.}~\bibnamefont {Zupan}},\ }\Doi
  {10.1103/PhysRevLett.107.012002} {\bibfield  {journal} {\bibinfo  {journal}
  {Phys.Rev.Lett.},\ }\textbf {\bibinfo {volume} {107}},\ \bibinfo {pages}
  {012002} (\bibinfo {year} {2011}{\natexlab{a}})},\ \Eprint
  {http://arxiv.org/abs/1102.3374} {arXiv:1102.3374 [hep-ph]} \BibitemShut
  {NoStop}%
\bibitem [{\citenamefont {Grinstein}\ \emph
  {et~al.}(2011){\natexlab{b}}\citenamefont {Grinstein}, \citenamefont {Kagan},
  \citenamefont {Zupan},\ and\ \citenamefont {Trott}}]{Grinstein:2011dz}%
  \BibitemOpen
  \bibfield  {author} {\bibinfo {author} {\bibfnamefont {B.}~\bibnamefont
  {Grinstein}}, \bibinfo {author} {\bibfnamefont {A.~L.}\ \bibnamefont
  {Kagan}}, \bibinfo {author} {\bibfnamefont {J.}~\bibnamefont {Zupan}}, \ and\
  \bibinfo {author} {\bibfnamefont {M.}~\bibnamefont {Trott}},\ }\Doi
  {10.1007/JHEP10(2011)072} {\bibfield  {journal} {\bibinfo  {journal} {JHEP},\
  }\textbf {\bibinfo {volume} {1110}},\ \bibinfo {pages} {072} (\bibinfo {year}
  {2011}{\natexlab{b}})},\ \Eprint {http://arxiv.org/abs/1108.4027}
  {arXiv:1108.4027 [hep-ph]} \BibitemShut {NoStop}%
\bibitem [{\citenamefont {Cheung}\ \emph {et~al.}(2009)\citenamefont {Cheung},
  \citenamefont {Keung},\ and\ \citenamefont {Yuan}}]{Cheung:2009ch}%
  \BibitemOpen
  \bibfield  {author} {\bibinfo {author} {\bibfnamefont {K.}~\bibnamefont
  {Cheung}}, \bibinfo {author} {\bibfnamefont {W.-Y.}\ \bibnamefont {Keung}}, \
  and\ \bibinfo {author} {\bibfnamefont {T.-C.}\ \bibnamefont {Yuan}},\ }\Doi
  {10.1016/j.physletb.2009.11.015} {\bibfield  {journal} {\bibinfo  {journal}
  {Phys.Lett.},\ }\textbf {\bibinfo {volume} {B682}},\ \bibinfo {pages} {287}
  (\bibinfo {year} {2009})},\ \Eprint {http://arxiv.org/abs/0908.2589}
  {arXiv:0908.2589 [hep-ph]} \BibitemShut {NoStop}%
\bibitem [{\citenamefont {Patel}\ and\ \citenamefont
  {Sharma}(2011)}]{Patel:2011eh}%
  \BibitemOpen
  \bibfield  {author} {\bibinfo {author} {\bibfnamefont {K.~M.}\ \bibnamefont
  {Patel}}\ and\ \bibinfo {author} {\bibfnamefont {P.}~\bibnamefont {Sharma}},\
  }\Doi {10.1007/JHEP04(2011)085} {\bibfield  {journal} {\bibinfo  {journal}
  {JHEP},\ }\textbf {\bibinfo {volume} {1104}},\ \bibinfo {pages} {085}
  (\bibinfo {year} {2011})},\ \Eprint {http://arxiv.org/abs/1102.4736}
  {arXiv:1102.4736 [hep-ph]} \BibitemShut {NoStop}%
\bibitem [{\citenamefont {Dimopoulos}\ \emph {et~al.}(1981)\citenamefont
  {Dimopoulos}, \citenamefont {Raby},\ and\ \citenamefont
  {Wilczek}}]{Dimopoulos:1981yj}%
  \BibitemOpen
  \bibfield  {author} {\bibinfo {author} {\bibfnamefont {S.}~\bibnamefont
  {Dimopoulos}}, \bibinfo {author} {\bibfnamefont {S.}~\bibnamefont {Raby}}, \
  and\ \bibinfo {author} {\bibfnamefont {F.}~\bibnamefont {Wilczek}},\ }\Doi
  {10.1103/PhysRevD.24.1681} {\bibfield  {journal} {\bibinfo  {journal}
  {Phys.Rev.},\ }\textbf {\bibinfo {volume} {D24}},\ \bibinfo {pages} {1681}
  (\bibinfo {year} {1981})}\BibitemShut {NoStop}%
\bibitem [{\citenamefont {Amaldi}\ \emph {et~al.}(1991)\citenamefont {Amaldi},
  \citenamefont {de~Boer},\ and\ \citenamefont {Furstenau}}]{Amaldi:1991cn}%
  \BibitemOpen
  \bibfield  {author} {\bibinfo {author} {\bibfnamefont {U.}~\bibnamefont
  {Amaldi}}, \bibinfo {author} {\bibfnamefont {W.}~\bibnamefont {de~Boer}}, \
  and\ \bibinfo {author} {\bibfnamefont {H.}~\bibnamefont {Furstenau}},\ }\Doi
  {10.1016/0370-2693(91)91641-8} {\bibfield  {journal} {\bibinfo  {journal}
  {Phys.Lett.},\ }\textbf {\bibinfo {volume} {B260}},\ \bibinfo {pages} {447}
  (\bibinfo {year} {1991})}\BibitemShut {NoStop}%
\bibitem [{\citenamefont {Aaltonen}\ \emph
  {et~al.}(2011){\natexlab{b}}\citenamefont {Aaltonen} \emph
  {et~al.}}]{Aaltonen:2010hza}%
  \BibitemOpen
  \bibfield  {author} {\bibinfo {author} {\bibfnamefont {T.}~\bibnamefont
  {Aaltonen}} \emph {et~al.} (\bibinfo {collaboration} {CDF Collaboration}),\
  }\Doi {10.1103/PhysRevD.83.071102} {\bibfield  {journal} {\bibinfo  {journal}
  {Phys.Rev.},\ }\textbf {\bibinfo {volume} {D83}},\ \bibinfo {pages} {071102}
  (\bibinfo {year} {2011}{\natexlab{b}})},\ \Eprint
  {http://arxiv.org/abs/1007.4423} {arXiv:1007.4423 [hep-ex]} \BibitemShut
  {NoStop}%
\bibitem [{\citenamefont {Martin}\ \emph {et~al.}(2009)\citenamefont {Martin},
  \citenamefont {Stirling}, \citenamefont {Thorne},\ and\ \citenamefont
  {Watt}}]{Martin:2009iq}%
  \BibitemOpen
  \bibfield  {author} {\bibinfo {author} {\bibfnamefont {A.}~\bibnamefont
  {Martin}}, \bibinfo {author} {\bibfnamefont {W.}~\bibnamefont {Stirling}},
  \bibinfo {author} {\bibfnamefont {R.}~\bibnamefont {Thorne}}, \ and\ \bibinfo
  {author} {\bibfnamefont {G.}~\bibnamefont {Watt}},\ }\Doi
  {10.1140/epjc/s10052-009-1072-5} {\bibfield  {journal} {\bibinfo  {journal}
  {Eur.Phys.J.},\ }\textbf {\bibinfo {volume} {C63}},\ \bibinfo {pages} {189}
  (\bibinfo {year} {2009})},\ \Eprint {http://arxiv.org/abs/0901.0002}
  {arXiv:0901.0002 [hep-ph]} \BibitemShut {NoStop}%
\bibitem [{\citenamefont {Han}\ \emph {et~al.}(2010)\citenamefont {Han},
  \citenamefont {Lewis},\ and\ \citenamefont {McElmurry}}]{Han:2009ya}%
  \BibitemOpen
  \bibfield  {author} {\bibinfo {author} {\bibfnamefont {T.}~\bibnamefont
  {Han}}, \bibinfo {author} {\bibfnamefont {I.}~\bibnamefont {Lewis}}, \ and\
  \bibinfo {author} {\bibfnamefont {T.}~\bibnamefont {McElmurry}},\ }\Doi
  {10.1007/JHEP01(2010)123} {\bibfield  {journal} {\bibinfo  {journal} {JHEP},\
  }\textbf {\bibinfo {volume} {1001}},\ \bibinfo {pages} {123} (\bibinfo {year}
  {2010})},\ \Eprint {http://arxiv.org/abs/0909.2666} {arXiv:0909.2666
  [hep-ph]} \BibitemShut {NoStop}%
\bibitem [{\citenamefont {Chen}\ \emph {et~al.}(2009)\citenamefont {Chen},
  \citenamefont {Klemm}, \citenamefont {Rentala},\ and\ \citenamefont
  {Wang}}]{Chen:2008hh}%
  \BibitemOpen
  \bibfield  {author} {\bibinfo {author} {\bibfnamefont {C.-R.}\ \bibnamefont
  {Chen}}, \bibinfo {author} {\bibfnamefont {W.}~\bibnamefont {Klemm}},
  \bibinfo {author} {\bibfnamefont {V.}~\bibnamefont {Rentala}}, \ and\
  \bibinfo {author} {\bibfnamefont {K.}~\bibnamefont {Wang}},\ }\Doi
  {10.1103/PhysRevD.79.054002} {\bibfield  {journal} {\bibinfo  {journal}
  {Phys.Rev.},\ }\textbf {\bibinfo {volume} {D79}},\ \bibinfo {pages} {054002}
  (\bibinfo {year} {2009})},\ \Eprint {http://arxiv.org/abs/0811.2105}
  {arXiv:0811.2105 [hep-ph]} \BibitemShut {NoStop}%
\end{thebibliography}%
\bibliographystyle{apsrev4-1}

\end{document}